\documentclass[11pt,amsmath,amssymb,noeprint]{revtex4-1}
\usepackage{amsfonts}
\usepackage{amsmath}
\usepackage{amssymb}
\usepackage{array}
\usepackage{bbding}
\usepackage{bm}
\usepackage{booktabs}
\usepackage{braket}
\usepackage{breqn}
\usepackage{color}

\usepackage{dcolumn}
\usepackage{diagbox}
\usepackage{endnotes}
\usepackage{epsfig}
\usepackage{epsf}
\usepackage{epstopdf}
\usepackage{extarrows}
\usepackage{graphicx}
\usepackage{graphics}
\usepackage{harpoon}
\usepackage{hyperref}
\usepackage{indentfirst}
\usepackage{makecell}
\usepackage{mathrsfs}
\usepackage{mathtools}
\usepackage{multirow}
\usepackage{pifont}
\usepackage{xcolor}

\usepackage{cleveref}

\newcommand{\RM}[1]{\textrm{\uppercase\expandafter{\romannumeral#1}}}

\begin{document}

\title{Relativistic correction to the binding energies of two-body hadronic molecular states}

\author{
Lin-Qing Song, and Hai-Qing Zhou\protect\footnotemark[1]\protect\footnotetext[1]{E-mail: zhouhq@seu.edu.cn} \\
School of Physics, Southeast University, Nanjing 211189, China
}
\date{\today}

\begin{abstract}

This study presents a systematic estimation of the relativistic correction to the binding energies of two-body hadronic molecular states by comparing the numerical solutions of Schr{\"o}dinger, Salpeter, and Bethe--Salpeter (BS) equations derived from the same one-boson-exchange (OBE) interaction, with the OBE input couplings kept fixed and without refitting the potential to each bound state. All BS results are computed in the ladder approximation. The numerical results reveal a counterintuitive, coupling-dependent property: for MeV-scale binding with heavy exchanged mesons, the binding energies from the ladder BS and Schr\"odinger equations at the same parameters can differ by large factors. \textcolor{black}{They show that shallow binding alone does not guarantee a small three-dimensional reduction effect when the interaction is short-ranged and strongly coupled.} Specifically, we first benchmark the effect in a scalar model and then extend the analysis to the $D\bar{D}$ system with hadronic form factors. In the shallow $D\bar{D}$ region illustrated, the magnitude of the ladder BS binding energy amounts to only $10\%\sim 30\%$ of the Schr\"odinger result at the same couplings. Such large discrepancies show that calculations based on the Schr\"odinger or Salpeter equations should be treated with caution when fixed OBE input parameters are transferred without matching or refitting. This conclusion does not conflict with the universality of shallow bound states, which concerns low-energy observables after short-distance parameters have been matched.

\end{abstract}

\maketitle
\section{Introduction}

Over the past two decades, a wide variety of exotic hadrons have been experimentally observed, including the $XYZ$, $P_c$, $P_{cs}$, and $T_{cc}^+$ states~\cite{Belle:2003nnu,BaBar:2005hhc,Belle:2007umv,Belle:2008xmh,BESIII:2013ris,BESIII:2013mhi,LHCb:2015yax,LHCb:2019kea,LHCb:2020jpq,LHCb:2021vvq}. Many of these states are not readily accommodated within the conventional quark model, and their internal structures have therefore become an important topic in hadronic physics. Various interpretations have been proposed, including compact tetraquarks and pentaquarks, hadronic molecules, hybrids, and kinematic effects~\cite{Lebed:2016hpi,Chen:2016qju,Guo:2017jvc,Liu:2019zoy,Brambilla:2019esw,Dong:2021juy,Wang:2025sic}. The proximity of several states, such as the $X(3872)$, $Y(4140)$, $Z_c(3900)$, $Z_c(4025)$, and $T_{cc}^+$, to nearby two-hadron thresholds has motivated extensive studies of their possible molecular nature~\cite{Lee:2009hy,Dong:2009yp,Ding:2009vj,Ding:2009vd,Ma:2014zva,Wang:2013cya,Guo:2013sya,Aceti:2014uea,He:2015mja,He:2013nwa,
Duan:2021pll,Li:2012ss,Liu:2019stu,Feijoo:2021ppq,Li:2021zbw,Lu:2025zae}. Similarly, the hidden-charm pentaquark states $P_c$ and $P_{cs}$ have been interpreted as molecular states composed of $\Sigma_c^{(*)}\bar{D}^{(*)}$, $\Xi_c^{(*)}\bar{D}^{(*)}$, and $\Xi_c^\prime\bar{D}^{(*)}$ pairs, as discussed in Refs.~\cite{Chen:2015loa,Shen:2016tzq,Chen:2016ryt,Xiao:2019gjd,Wang:2019nvm,Zhu:2021lhd,Zou:2021sha,Feijoo:2022rxf,Xiao:2022csb}.

Theoretically, the Bethe--Salpeter (BS) equation serves as an essential tool to investigate relativistic bound-state problems within the framework of quantum field theory~\cite{Salpeter:1951sz,Gell-Mann:1951ooy}. However, due to the complexity of the formalism, theoretical studies of hadronic molecular states often resort to three-dimensional (3D) reductions. Common reductions include the instantaneous approximation~\cite{Salpeter:1952ib,Sazdjian:1985pg,Sazdjian:1986aw}, the quasipotential approximation~\cite{Todorov:1970gr,Connell:1991ae}, and the nonrelativistic (NR) approximation~\cite{PhysRev.100.912,PhysRev.104.1782}. Consequently, most existing literature relies on the Schr\"{o}dinger equation with a one-boson-exchange (OBE) potential, reduced BS equations, or unitary approaches. In this study, we systematically compare the binding energies of two-body hadronic molecular states obtained from the BS, Salpeter, and Schr{\"o}dinger equations using the same underlying interaction and fixed OBE input parameters. In this comparison, the BS--Schr\"odinger difference reflects the relativistic effects neglected in the reduction, while the Salpeter result is used to separate the instantaneous and NR steps.

\section{Formalism}

We first introduce the formalism using a massive-exchange Wick--Cutkosky-type scalar model. The system consists of two identical scalar particles of mass $m$ interacting through the exchange of a scalar boson with mass $m_{ex}$. The interaction vertex is defined by $-2igm$, where the coupling constant $g$ is dimensionless.

The ladder BS equation for the two-body $T$-matrix in this model can be written as
\begin{equation}
T(P,k,k') = K_{\mathrm{BS}}^{\mathrm{OBE}}(k-k') + \int \frac{d^4 q}{(2\pi)^4} K_{\mathrm{BS}}^{\mathrm{OBE}}(k-q) G(P,q) T(P,q,k'),
\label{Eq:BS-equation-for-T}
\end{equation}
where $P$ is the total four-momentum of the system, $k, k'$, and $q$ are the relative four-momenta, and
\begin{align}
K_{\mathrm{BS}}^{\mathrm{OBE}}(k-q) &= \frac{-4ig^2m^2}{(k-q)^2-m_{ex}^2+i\epsilon}, \notag \\
G(P, q) &= \frac{-1}{(q_1^2-m^2+i \epsilon)(q_2^2-m^2+i \epsilon)},
\label{equation:4D-BS-equation-potential}
\end{align}
with $k_{1,2}=\frac{1}{2}P\pm k, q_{1,2}=\frac{1}{2}P\pm q$ denoting the momenta of the constituent particles.

If the system has a $J=0$ bound state of mass $M$, the scattering amplitude has a pole as $P^2\to M^2$:
\begin{equation}
T^{J=0}(P,k, k') = \frac{\chi(P,k) \chi^\dagger(P,k')}{P^2 - M^2} + T_{\text{reg}}(P, k, k'),
\label{Eq:T-matrix-to-BS-wave}
\end{equation}
where $\chi(P,k)$ is the BS amplitude for a $J=0$ state and $T_{\text{reg}}$ is the regular part of the amplitude, which is analytic at the pole.

Substituting Eq.~(\ref{Eq:T-matrix-to-BS-wave}) into Eq.~(\ref{Eq:BS-equation-for-T}) and taking the limit $P^2\rightarrow M^2$, one obtains the homogeneous BS equation for the BS amplitude:
\begin{align}
\chi_{\mathrm{BS}}(P,k) = \int \frac{d^4q}{(2\pi)^4}K_{\mathrm{BS}}^{\mathrm{OBE}}(k-q)G(P,q)\chi_{\mathrm{BS}}(P,q).
\label{equation:4D-BS-general-1}
\end{align}

By defining $P \equiv (M,\boldsymbol{0})$, $k \equiv(k_0,\boldsymbol{k})$, and $q \equiv(q_0,\boldsymbol{q})$, Eq.~(\ref{equation:4D-BS-general-1}) can be written as
\begin{equation}
\chi_{\text{BS}}(M,k_0,|\boldsymbol{k}|) = \int \frac{dq_0 d|\boldsymbol{q}|}{(2\pi)^4} |\boldsymbol{q}|^2 \left[\int d\Omega_{\boldsymbol{q}}K_{\text{BS}}^{\text{OBE}}(k-q)\right] G(M,q_0,|\boldsymbol{q}|)\chi_{\text{BS}}(M,q_0,|\boldsymbol{q}|).
\label{equation:2D-BS-equation-general}
\end{equation}
Approaches to solving such an equation (or its equivalent forms) have been extensively investigated in Refs.~\cite{Wick:1954eu,Nieuwenhuis:1996qx,Efimov:2003hs,Karmanov:2005nv,Carbonell:2010tz}. In the present work, we employ the Wick rotation method on the BS equation and solve it in Euclidean space.

{\color{black}
For fixed $|\boldsymbol q|$, the two constituent propagators in
$G(P,q)$ give four poles in the complex $q_0$ plane as
\begin{align}
q_0^{(1)}&=-\frac{M}{2}+\omega_q-i\epsilon, \nonumber\\
q_0^{(2)}&=-\frac{M}{2}-\omega_q+i\epsilon, \nonumber\\
q_0^{(3)}&= \frac{M}{2}-\omega_q+i\epsilon, \nonumber\\
q_0^{(4)}&= \frac{M}{2}+\omega_q-i\epsilon,
\end{align}
with $\omega_q=\sqrt{m^2+|\boldsymbol q|^2}$.
The exchanged-boson propagator in $K_{\mathrm{BS}}^{\mathrm{OBE}}(k-q)$ gives two
additional poles as
\begin{equation}
q_0^{(\pm)}
=
k_0\pm\sqrt{\Omega_{\boldsymbol{kq}}^2-i\epsilon},
\end{equation}
with  $\Omega_{\boldsymbol{kq}}^2
=
(\boldsymbol k-\boldsymbol q)^2+m_{ex}^2$.

In the present calculation, the Euclidean BS equation is obtained by
analytically continuing the full Minkowski BS equation through a continuous
deformation of both the external relative energy and the internal
relative-energy integration contour. The continuation can be parameterized as
\begin{equation}
k_0=e^{i\theta}\bar{k}_0,
\end{equation}
with $0\leq\theta\leq\frac{\pi}{2}$.  The $q_0$ integration contour is defined as
\begin{equation}
C_\theta:
\qquad
q_0=e^{i\theta}\bar{q}_0,
\end{equation}
with  $-\infty<\bar q_0<\infty $.

For a bound state with $M<2m$, one has $\omega_q\geq m>\frac{M}{2}$.
The four constituent poles therefore lie in the second and fourth quadrants
of the complex $q_0$ plane, and none of them is crossed when the
contour is continuously rotated from $C_0$ to $C_{\pi/2}$.

The denominator of the exchanged-boson propagator on the contour is
\begin{align}
D_{ex}(\theta)
&=
e^{2i\theta}\Delta^2
-\Omega_{\boldsymbol{kq}}^2+i\epsilon
=
\left[
\Delta^2\cos(2\theta)-\Omega_{\boldsymbol{kq}}^2
\right]
+i\left[
\Delta^2\sin(2\theta)+\epsilon
\right],
\end{align}
with  $\Delta=\bar k_0-\bar q_0$. Its real and imaginary parts cannot vanish simultaneously for $0<\theta<\pi/2$. At the Euclidean endpoint,
\begin{align*}
D_{ex}\left(\frac{\pi}{2}\right)
=-\left[(\bar{k}_0-\bar{q}_0)^2
+(\boldsymbol{k}-\boldsymbol{q})^2+m_{ex}^2\right]+i\epsilon,
\end{align*}
which is nonsingular for $m_{ex}>0$. Therefore, none of these explicit poles crosses the continuously deformed contour, and the
$q_0$ integration path can be rotated from the real axis to the imaginary
axis while $k_0$ is analytically continued at the same time. Consequently,
no additional residue or principal-value term is generated during the Wick
rotation, and the Minkowski BS equation is analytically continued to the
corresponding Euclidean BS equation. Direct calculations in Minkowski space
also reproduce the same bound-state spectrum as the Euclidean formulation
for the massive OBE interaction~\cite{Karmanov:2005nv}.
}

After performing Wick rotation and the angular integration, Eq.~(\ref{equation:2D-BS-equation-general}) can be written as
\begin{align}
\chi_{\text{BS}}(M,\bar{k}_0,|\boldsymbol{k}|) &= i\int \frac{d\bar{q}_0 d|\boldsymbol{q}|}{(2\pi)^4} |\boldsymbol{q}|^2 K^{(0)}_{ \text{BS}}(\bar{k}_0,\bar{q}_0,|\boldsymbol{k}|,|\boldsymbol{q}|)G(M,i\bar{q}_0,|\boldsymbol{q}|)\chi_{\text{BS}}(M,\bar{q}_0,|\boldsymbol{q}|),
\label{equation:2D-BS-equation-analytical}
\end{align}
where
\begin{align}
K^{(0)}_{\text{BS}}(\bar{k}_0,\bar{q}_0,|\boldsymbol{k}|,|\boldsymbol{q}|)&\equiv \int d\Omega_{\boldsymbol{q}} K_{\text{BS}}^{\text{OBE}}(k-q)  =\frac{4ig^2 m^{2}\pi}{|\boldsymbol{k}||\boldsymbol{q}|} \log \frac{(|\boldsymbol{k}|+|\boldsymbol{q}|)^2+m_{ex}^2+(\bar{k}_{0} -  \bar{q}_{0})^2}{(|\boldsymbol{k}|-|\boldsymbol{q}|)^2+m_{ex}^2+(\bar{k}_{0} -  \bar{q}_{0})^2}.
\label{equation:2D-BS-equation-potential}
\end{align}

In the literature, the instantaneous approximation is commonly applied as~\cite{Kopaleishvili:2001cr,Chang:2004im,Lucha:2005nz}
\begin{align}
K_{\text{BS}}^{\text{OBE}}(k-q) &\approx  K_{\text{S}}^{\text{OBE}}(\boldsymbol{k}-\boldsymbol{q}) =\frac{4ig^2 m^{2}}{(\boldsymbol{k}-\boldsymbol{q})^2+m_{ex}^2-i\epsilon}.
\label{equation:K-instantaneous-approximation}
\end{align}
After integrating over the variable $q_0$, Eq.~(\ref{equation:2D-BS-equation-general}) simplifies to the one-dimensional (1D) form of the Salpeter equation as
\begin{align}
\chi_{\text{S}}(M,|\boldsymbol{k}|) = \int\frac{d|\boldsymbol{q}|}{(2\pi)^4}|\boldsymbol{q}|^{2}  K^{(0)}_{\text{S}}(|\boldsymbol{k}|,|\boldsymbol{q}|)\bar{G}(M,|\boldsymbol{q}|)\chi_{\text{S}}(M,|\boldsymbol{q}|),
\label{equation:1D-Salpeter-equation}
\end{align}
where
\begin{align}
K^{(0)}_{\text{S}}(|\boldsymbol{k}|,|\boldsymbol{q}|)&= K^{(0)}_{\text{BS}}(0,0,|\boldsymbol{k}|,|\boldsymbol{q}|),\notag\\
\bar{G}(M,|\boldsymbol{q}|) &\equiv \int dq_0 G(M,q_0,|\boldsymbol{q}|)
=\frac{2i\pi}{\omega_q}\frac{1}{M^2-4\omega_q^2+i\epsilon} \nonumber\\
&=\frac{i\pi}{2\omega_q^2}
\left[
\frac{1}{M-2\omega_q+i\epsilon}
-\frac{1}{M+2\omega_q-i\epsilon}
\right],
\label{equation:1D-Salpeter-equation-propagator}
\end{align}
with $\omega_q=\sqrt{m^2+|\boldsymbol{q}|^2}$. {\color{black}
The two terms in Eq.~(\ref{equation:1D-Salpeter-equation-propagator}) are the positive- and negative-energy contributions, respectively. In the NR approximation, the negative-energy term is omitted, while the positive-energy term and its prefactor are expanded under the conditions $|\boldsymbol{q}|/m\ll 1$, $|\boldsymbol{k}|/m\ll 1$, and $|E|/m\ll 1$, where $E\equiv M-2m$ is negative for a bound state. One then obtains
}
\begin{align}
\bar{G}(M,|\boldsymbol{q}|)\approx \bar{G}_{\text{NR}}(M,|\boldsymbol{q}|) = \frac{i\pi}{2m^2}\frac{1}{E-|\boldsymbol{q}|^2/(2\mu)+i\epsilon},
\end{align}
with reduced mass $\mu =m/2$.

Consequently, one obtains a Schr{\"o}dinger-like equation as
\begin{align}
\chi_{\text{Sch}}(|\boldsymbol{k}|) &= \int\frac{d|\boldsymbol{q}|}{(2\pi)^4}|\boldsymbol{q}|^2
K^{(0)}_{\text{S}}(|\boldsymbol{k}|,|\boldsymbol{q}|)\bar{G}_{\text{NR}}(M,|\boldsymbol{q}|) \chi_{\text{Sch}}(|\boldsymbol{q}|).
\label{equation:1D-Schrodinger-equation}
\end{align}

Since $E<0$ for a bound state, $\bar{G}_{\text{NR}}(M,|\boldsymbol{q}|)$ has no pole for real $|\boldsymbol{q}|$. Defining
\begin{align}
\phi_{\text{Sch}}(|\boldsymbol{q}|) \equiv \bar{G}_{\text{NR}}(M, |\boldsymbol{q}|) \chi_{\text{Sch}}(|\boldsymbol{q}|),
\end{align}
Eq.~(\ref{equation:1D-Schrodinger-equation}) can be rewritten as
\begin{align}
\left(E - \frac{|\boldsymbol{k}|^2}{2\mu} + i\epsilon\right) \phi_{\text{Sch}}(|\boldsymbol{k}|) &= \frac{i\pi}{2m^2}\int \frac{d|\boldsymbol{q}|}{(2\pi)^4} |\boldsymbol{q}|^2 K^{(0)}_{\text{S}}(|\boldsymbol{k}|, |\boldsymbol{q}|) \phi_{\text{Sch}}(|\boldsymbol{q}|),
\end{align}
where the specific form of $\bar{G}_{\text{NR}}(M, |\boldsymbol{k}|)$ has been used. This derivation can also be generalized to higher partial waves. With the standard three-dimensional normalization, the corresponding equation takes the form
\begin{align}
\left(E - \frac{|\boldsymbol{k}|^2}{2\mu} + i\epsilon\right) \phi_{\text{Sch}}(\boldsymbol{k}) &=  \int \frac{d^3\boldsymbol{q}}{(2\pi)^3} \tilde{K}(\boldsymbol{k}-\boldsymbol{q}) \phi_{\text{Sch}}(\boldsymbol{q}),
\end{align}
where the real static kernel is
\begin{align}
\tilde{K}(\boldsymbol{p}) &\equiv \frac{i}{4m^2}K^{\text{OBE}}_{\text{S}}(\boldsymbol{p})
=-\frac{g^2}{|\boldsymbol{p}|^2+m_{ex}^2}.
\end{align}
Defining
\begin{align}
\phi_{\text{Sch}}(\boldsymbol{k}) &\equiv \int d^3\boldsymbol{r} e^{-i\boldsymbol{k}\cdot\boldsymbol{r}} \psi_{\text{Sch}}(\boldsymbol{r}),
\end{align}
one obtains
\begin{align}
\left[E + \frac{\nabla^2}{2\mu}\right] \psi_{\text{Sch}}(\boldsymbol{r}) &= V(\boldsymbol{r}) \psi_{\text{Sch}}(\boldsymbol{r}),
\end{align}
which is precisely the usual Schr\"{o}dinger equation, where
\begin{align}
V(\boldsymbol{r}) &\equiv \int \frac{d^3\boldsymbol{q}}{(2\pi)^3} e^{i\boldsymbol{q}\cdot\boldsymbol{r}} \tilde{K}(\boldsymbol{q})
=-\alpha\frac{e^{-m_{ex}r}}{r}.
\end{align}
The above steps also show that $\bar{G}_{\text{NR}}$ has poles on the real momentum axis above threshold. Therefore, the bound-state numerical procedure used here cannot be applied directly to scattering states without treating these poles and imposing the appropriate scattering boundary conditions.

We emphasize that Eqs.~(\ref{equation:2D-BS-equation-analytical},~\ref{equation:1D-Salpeter-equation},~\ref{equation:1D-Schrodinger-equation}) are derived from the same underlying interaction. Thus, their differences arise solely from the approximations employed.

\section{Numerical Method}

A standard eigenvalue form for Eqs.~(\ref{equation:2D-BS-equation-analytical},~\ref{equation:1D-Salpeter-equation},~\ref{equation:1D-Schrodinger-equation}) is
\begin{align}
\chi(x) &= \lambda(E)\,[\hat{X}(E)\chi](x),
\end{align}
where $\hat{X}(E)$ denotes the integral operator corresponding to Eqs.~(\ref{equation:2D-BS-equation-analytical},~\ref{equation:1D-Salpeter-equation},~\ref{equation:1D-Schrodinger-equation}). Here, $x=|\boldsymbol{k}|$ for the Schr\"{o}dinger and Salpeter equations, while $x=(\bar{k}_0,|\boldsymbol{k}|)$ for the Euclidean BS equation. For a fixed energy $E$, the physical solution is obtained when one of the eigenvalues $\lambda(E)$ equals $1$. In the present calculation, we instead rewrite the equations as
\begin{align}
\chi(x) &= \eta(E)+[\hat{X}(E)\chi](x).
\label{eq:general-form-with-eta}
\end{align}
By imposing the normalization $\chi(x_0)=1$ at a chosen point $x_0$, one obtains a system of linear equations for $\eta$ and $\chi$ at a given $E$. The zeros of $\eta(E)$ give the bound-state energies, and multiple zeros correspond to multiple bound states.

In our practical implementation, Eq.~(\ref{equation:2D-BS-equation-analytical}) is first recast into the form of Eq.~(\ref{eq:general-form-with-eta}) under the normalization condition $\chi(0,0)=1$. We employ the Gauss--Legendre quadrature rule for the integral variables, which are initially defined on the unit interval $[0,1]$. A tangent-based transformation is then applied to map these nodes to finite numerical integration domains: $[0,q_{\max}]$ for the magnitudes $|\bm q|$ and $|\bm k|$, and $[-q_{\max},q_{\max}]$ for the temporal components $\bar{q}_0$ and $\bar{k}_0$. Specifically, for the momentum magnitude $|\bm{q}|$, the mapping is defined as
\begin{align}
|\boldsymbol{q}|_i &\equiv \tan[x_i\arctan(q_{\max})],\nonumber \\
 w_i &\equiv J_i \bar{w}_i,
\label{eq:map2}
\end{align}
where $x_i$ and $\bar{w}_i$ are the nodes and weights of the Gauss--Legendre quadrature, respectively, and $J_i$ denotes the corresponding Jacobian factor.

By discretizing $\bar{q}_0$ and $|\bm{q}|$ on an $800 \times 800$ grid, we obtain a system of $6.4 \times 10^5$ inhomogeneous linear equations for $\eta$ and the wave function at a given energy $E$. These equations are solved using the PETSc package~\cite{PETSc:2026}. Once $\eta(E)$ is determined, the bisection method is employed to find the physical energy $E$ that satisfies the bound-state condition $\eta(E)=0$.

{\color{black}
In the numerical implementation, $q_{\max}$ is the finite endpoint used to approximate the infinite integration domain. It is not a renormalization scale or a regulator in effective field theory (EFT). The agreement of the binding energies for $q_{\max}=10$, $100$, and $1000$ GeV is therefore a numerical convergence test. The converged BS binding energies agree with Ref.~\cite{Karmanov:2005nv}. The numerical solutions for Eqs.~(\ref{equation:1D-Salpeter-equation}) and (\ref{equation:1D-Schrodinger-equation}) can be obtained in the same manner.
}

\section{Numerical Results and Discussion}

\subsection{Scalar system}

\begin{figure}[htbp]
  \centering
  \begin{minipage}[t]{0.48\textwidth}
    \centering
    \includegraphics[width=\linewidth]{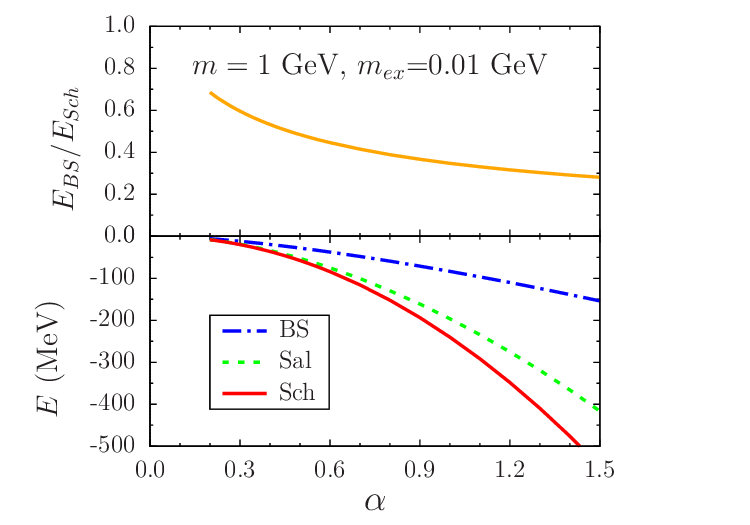}
  \end{minipage}
  \hspace{0.02\textwidth}
  \begin{minipage}[t]{0.48\textwidth}
    \centering
    \includegraphics[width=\linewidth]{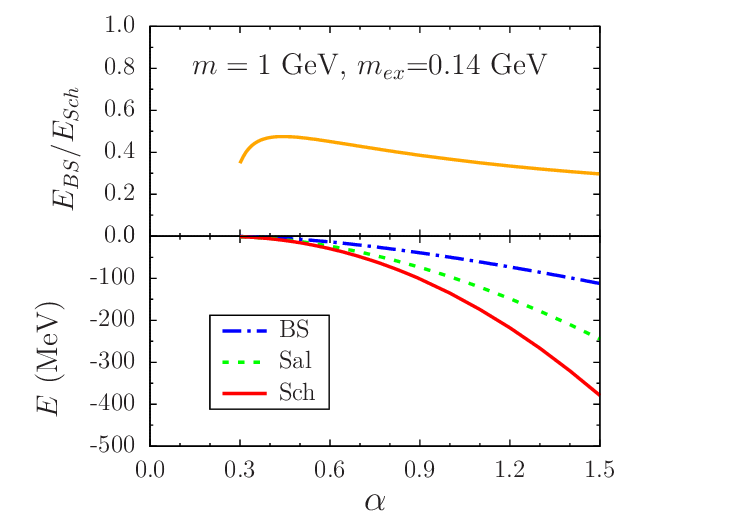}
  \end{minipage}

  \vspace{0.35em}

  \begin{minipage}[t]{0.48\textwidth}
    \centering
    \includegraphics[width=\linewidth]{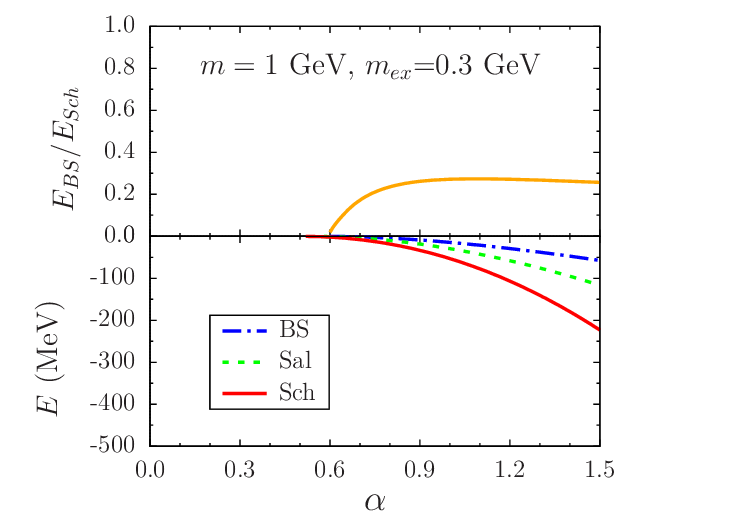}
  \end{minipage}
  \hspace{0.02\textwidth}
  \begin{minipage}[t]{0.48\textwidth}
    \centering
    \includegraphics[width=\linewidth]{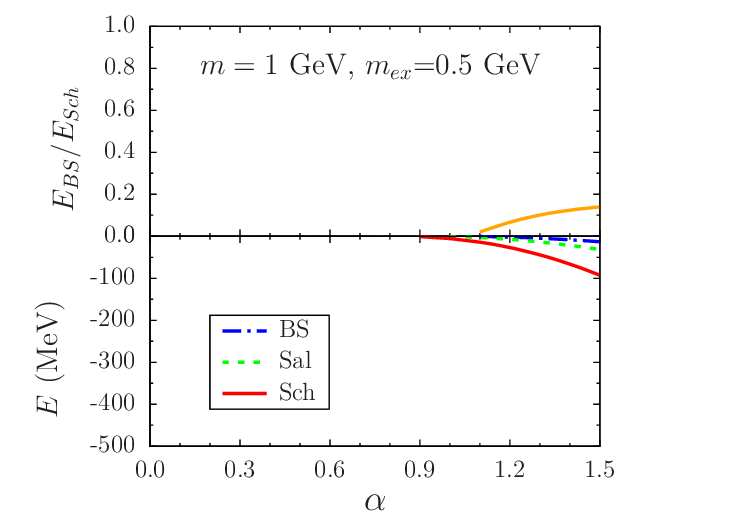}
  \end{minipage}
  \caption{Ground-state binding energy $E$ for $m_{ex}=0.01$, $0.14$, $0.3$, and $0.5$ GeV as a function of $\alpha$. The blue dash-dot, green dot, and red solid curves correspond to the BS, Salpeter, and Schr{\"o}dinger equations, respectively. The orange curve is $E_{\rm BS}/E_{\rm Sch}$ in the common bound-state region.}
\label{Fig:Eb-alpha-m=1}
\end{figure}

{\color{black}
In Fig.~\ref{Fig:Eb-alpha-m=1}, we show the binding energies as functions of $\alpha$ (with $\alpha\equiv g^2/(4\pi)$) for $m=1$ GeV. The four values $m_{ex}=0.01$, $0.14$, $0.3$, and $0.5$ GeV are used to study the dependence on the interaction range, and are not all identified with particular exchanged hadrons. The value $m_{ex}=0.01$ GeV is retained as a long-range theoretical benchmark, $m_{ex}=0.14$ GeV is close to the pion mass scale, and $m_{ex}=0.3$ and $0.5$ GeV represent shorter-range interactions.

For nonzero $m_{ex}$, the Schr\"odinger and BS equations each have a critical coupling at which the first bound state appears. We denote these critical couplings by $\alpha_c^{\rm Sch}$ and $\alpha_c^{\rm BS}$, respectively. When $\alpha_c^{\rm Sch}<\alpha_c^{\rm BS}$, the ratio $E_{\rm BS}/E_{\rm Sch}$ vanishes as $\alpha$ approaches $\alpha_c^{\rm BS}$ from above, because $E_{\rm BS}$ approaches zero while $E_{\rm Sch}$ is already finite. For the smallest exchange mass, the two critical couplings are close, so this initial rise occurs in a narrow interval near threshold; over the range displayed, the ratio is already decreasing as the coupling increases. For larger exchange masses, the separation of the critical couplings becomes more pronounced and the initial increase is visible. At stronger coupling, the growing difference between the BS and Schr\"odinger results tends to reduce the ratio. The competition between these effects explains why the orange curve for an intermediate exchange mass first increases and then decreases, while the heavier-exchange curve is still increasing over the range shown.

At $|E_{\rm Sch}|\simeq 20$ MeV, $E_{\rm BS}/E_{\rm Sch}$ is about $60\%$ for $m_{ex}=0.01$ GeV and about $5\%$ for $m_{ex}=0.5$ GeV. This comparison shows that the difference between the two equations depends not only on the binding energy, but also strongly on the exchange mass.
}

To understand why the three equations give different binding energies, we next examine the reductions in more detail. The BS equation for the $T$-matrix in operator form can be written as
\begin{equation}
T = K_{\mathrm{BS}}^{\mathrm{OBE}} + K_{\mathrm{BS}}^{\mathrm{OBE}} G T.
\label{BS-eq-original}
\end{equation}
An auxiliary kernel $\bar{K}$ and a reduced propagator $G_{\rm red}$ can be introduced such that the following relation remains valid~\cite{Lepage:1977gd,Caswell:1978mt}:
\begin{equation}
T = \bar{K} + \bar{K} G_{\rm red} T.
\label{BS-eq-with-Kbar}
\end{equation}
By comparing Eqs.~(\ref{BS-eq-original}) and (\ref{BS-eq-with-Kbar}), one obtains the exact diagrammatic series for the new kernel:
\begin{equation}
\begin{aligned}
\bar{K} &= [I - K_{\mathrm{BS}}^{\mathrm{OBE}}(G-G_{\rm red})]^{-1} K_{\mathrm{BS}}^{\mathrm{OBE}} \\
&= K_{\mathrm{BS}}^{\mathrm{OBE}} + K_{\mathrm{BS}}^{\mathrm{OBE}}(G-G_{\rm red}) K_{\mathrm{BS}}^{\mathrm{OBE}} + \cdots,
\end{aligned}
\end{equation}
where the second equality follows from the geometric series expansion. This relation is mathematically exact provided that the operator is invertible.

Based on this expansion, various approximation schemes can be employed. For instance, by setting $G_{\rm red}=\bar{G}_{\text{NR}}(M,|\boldsymbol{q}|)\delta(q_0)$, the BS equation for the amplitude $\chi_{\mathrm{BS}}$ (for a $J=0$ state) can be expressed as
\begin{equation}
\chi_{\mathrm{BS}}(M, k) = \int \frac{d^4q}{(2\pi)^4} \bar{K}(M, k, q) \bar{G}_{\text{NR}}(M, \boldsymbol{q}) \delta(q_0) \chi_{\mathrm{BS}}(M, q).
\label{Eq:BS-expansion-pre}
\end{equation}
By taking the instantaneous limit $k_0 = 0$, one has:
\begin{equation}
\begin{aligned}
\chi_{\mathrm{BS}}(M, 0, |\boldsymbol{k}|) &= \int \frac{d^3\boldsymbol{q}}{(2\pi)^4} K_{\mathrm{S}}^{\mathrm{OBE}}(\boldsymbol{k}-\boldsymbol{q}) \bar{G}_{\text{NR}}(M, \boldsymbol{q}) \chi_{\mathrm{BS}}(M, 0, |\boldsymbol{q}|) \\
& \quad + \int \frac{d^3\boldsymbol{q}}{(2\pi)^4} \delta K(M, \boldsymbol{k}, \boldsymbol{q}) \bar{G}_{\text{NR}}(M, \boldsymbol{q}) \chi_{\mathrm{BS}}(M, 0, |\boldsymbol{q}|),
\end{aligned}
\label{Eq:BS-expansion}
\end{equation}
where the correction term is defined as:
\begin{equation}
\delta K(M, \boldsymbol{k}, \boldsymbol{q}) = K_{\mathrm{BS}}^{\mathrm{OBE}}(G-G_{\rm red}) K_{\mathrm{BS}}^{\mathrm{OBE}} \Big|_{k_0=q_0=0} + \cdots
\end{equation}

The standard Schr\"{o}dinger equation is recovered by retaining only the first integral in Eq.~(\ref{Eq:BS-expansion}) and neglecting the contribution from $\delta K$. In QED bound-state systems, the small fine-structure constant $\alpha_{\mathrm{QED}}$ suppresses the higher-order terms in $\delta K$, making the Schr\"{o}dinger equation a highly accurate leading approximation~\cite{Lepage:1977gd,Caswell:1978mt}. In the Coulomb gauge, the leading Coulomb interaction is instantaneous, while the remaining contributions enter at higher orders.

In the systems studied here, a much larger coupling is needed to produce a bound state for a given exchange mass. The numerical differences above show that the terms in $\delta K$ are not always small. Thus, the accuracy of the instantaneous and nonrelativistic approximations depends on both the propagator difference $G-G_{\rm red}$ and the coupling $\alpha$ in $K_{\mathrm{BS}}$. This differs from hydrogen-like systems, where the small coupling makes these corrections small.

\begin{table}[htbp]
\centering
\begin{tabular}{c c c c c c}
\hline
$\alpha=1.5$ & $q_{\max}=1000$ & $q_{\max}=100$ & $q_{\max}=10$ & $q_{\max}=1$ & $q_{\max}=0.6$ \\
\hline
BS            & -0.012952 & -0.012952 & -0.012951 & -0.008553 & -0.002899 \\
Salpeter      & -0.031474 & -0.031474 & -0.031470 & -0.018261 & -0.005682 \\
Schr\"odinger & -0.092753 & -0.092753 & -0.092316 & -0.030064 & -0.007845 \\
\hline
\end{tabular}
\caption{Binding energy $E$ calculated from the ladder BS, Salpeter, and Schr\"odinger equations for several numerical integration endpoints $q_{\max}$, with $m=1$ GeV, $m_{ex}=0.5$ GeV, and $\alpha=1.5$. All results and endpoint values are in units of GeV.}
\label{tab:data-alpha-1.5}
\end{table}

{\color{black}
Tab.~\ref{tab:data-alpha-1.5} shows the dependence on the numerical integration endpoint $q_{\max}$. In Tab.~\ref{tab:data-alpha-1.5}, increasing the numerical integration endpoint from $q_{\max}=1$ to $10$ GeV changes the absolute values of the BS, Salpeter, and Schr\"odinger binding energies from $0.008553$, $0.018261$, and $0.030064$ GeV to $0.012951$, $0.031470$, and $0.092316$ GeV, respectively. At $q_{\max}=1$ GeV, the results retain about $66\%$, $58\%$, and $33\%$ of their respective values at $q_{\max}=10$ GeV. The Schr\"odinger result is most sensitive to including momenta above $1$ GeV, while the BS result is least sensitive. Changing $q_{\max}$ changes the full eigenvalue problem, so these differences are not additive contributions from separate momentum bins. In the BS equation, $q_{\max}$ limits both the Euclidean relative energy and the three-momentum. In the Salpeter and Schr\"odinger equations, it limits only the three-momentum. These results show that momenta outside the strict nonrelativistic region can be important even in the reduced Schr\"odinger equation. This is only a momentum-domain diagnostic within the single-channel equations used here. It does not determine the contributions from additional scattering channels, which depend on their threshold separations and transition couplings and require a separate coupled-channel calculation. The results are stable when $q_{\max}$ increases from $10$ to $1000$ GeV, so the sensitivity comes from the intermediate- and high-momentum region rather than the very high-momentum region.
}

Fig.~\ref{Fig:Eb-alpha-m=1} and Tab.~\ref{tab:data-alpha-1.5} show large differences among the three equations when the same OBE parameters are used without refitting. Nevertheless, the nonrelativistic Schr\"odinger equation often provides a useful description of shallow bound states in hadronic and nuclear physics. For $|E|\ll m$, the inverse binding momentum, $R_0=1/\sqrt{2\mu|E|}$, provides a characteristic length scale of the bound state. In the universal regime with a large scattering length, low-energy observables can be described in terms of the scattering length and effective range and, after matching, are insensitive to the detailed short-range dynamics~\cite{Braaten:2004rn}. Our results do not contradict this EFT picture. EFT universality links physical observables after short-distance physics has been integrated out. We instead examine whether the same fixed OBE input parameters, when carried through the ladder BS, instantaneous Salpeter, and NR Schr\"odinger reductions of one kernel, yield consistent binding energies without parameter matching or refitting.

In typical phenomenology the Schr\"odinger potential is an effective interaction. Couplings and ranges are fitted to phase shifts, binding energies, or other data. Retardation, ladder truncation, and higher-order effects may then be absorbed into the fitted parameters. Experimental agreement does not require the NR reduction to be correct at fixed OBE input couplings. It requires only a good effective potential for the observables studied. Our large discrepancies show a different point. The same fixed OBE input parameters should not be used blindly in a ladder BS equation and in a reduced Schr\"odinger equation. Parameter matching or phenomenological refitting is needed.

{\color{black}
The comparison above concerns the instantaneous and NR reductions of one common ladder OBE kernel. Corrections to the BS kernel are not included in this comparison and constitute a separate effect. They can also be large. For the same scalar model with $m_{ex}/m=0.5$, Ref.~\cite{Carbonell:2005cl} gives $-E_{\rm BS}/m=0.10$ for the ladder kernel at $\alpha=2.50$. When the crossed-ladder contribution is included, $-E_{\rm BS}/m=0.20$ is already reached at $\alpha=2.42$. Since the magnitude of the binding energy increases with $\alpha$, the ladder-plus-crossed-ladder result at $\alpha=2.50$ must be larger than $0.20$. Thus, in this strong-coupling example, the crossed-ladder contribution more than doubles the magnitude of the ladder binding energy. Generalized-ladder calculations in the Feynman--Schwinger representation show the same qualitative behavior~\cite{Nieuwenhuis:1996pl}. Related crossed-boson calculations in the BS framework also find large corrections~\cite{Theussl:1999xq}.

These results set the scope of the present work. The ladder BS equation is used here as the four-dimensional reference for examining the instantaneous and NR reductions at fixed OBE parameters. It is not regarded as a complete physical prediction. Beyond-ladder corrections, together with the coupled-channel effects, are outside the present comparison and may produce additional changes.
}

\subsection{\texorpdfstring{$D\bar{D}$}{D-anti-D} system}

{\color{black}
We next apply the same comparison to the isoscalar $S$-wave $D\bar{D}$ system with $J^{PC}=0^{++}$ considered in Ref.~\cite{Wang:2021aql}. We use the same OBE parameters as inputs and compare the results obtained from the BS, Salpeter, and Schr\"odinger equations. The purpose is to examine the reduction effect for a hadronic OBE interaction, rather than to give a complete prediction for the $D\bar{D}$ system. The interaction includes $\rho$, $\omega$, and $\sigma$ exchanges. The exchanged particles and vertex structures differ from those of the scalar model, but the interaction has a similar OBE form and can be used for the same comparison. After including the hadronic form factors, the kernels are replaced by
}
\begin{align}
K^{\text{OBE}}_{\text{BS}}(k-q) &\rightarrow K^{D\bar{D}}_{\text{BS}}(k-q) = \sum_{i=\omega,\rho,\sigma}K^{i}_{\text{BS}}(k-q) F_i(k-q), \nonumber\\
K^{\text{OBE}}_{\text{S}}(\boldsymbol{k}-\boldsymbol{q}) &\rightarrow K^{D\bar{D}}_{\text{S}}(\boldsymbol{k}-\boldsymbol{q}) = \sum_{i=\omega,\rho,\sigma}K^{i}_{ \text{S}}(\boldsymbol{k}-\boldsymbol{q})F_i(k-q)\Big|_{k_0-q_0=0},
\label{equation:BS-potential-dipole}
\end{align}
where the product of the two monopole vertex form factors is denoted by
\begin{align}
F_i(k-q) = \left( \frac{\Lambda^2-m_{i}^2}{(k-q)^2 - \Lambda^2} \right)^2,
\label{equation:form-factor-dipole}
\end{align}
and $K^{i}_{\text{BS,S}}$ corresponds to $K^{\text{OBE}}_{\text{BS,S}}$ with the corresponding $m_i$ and $\alpha_i$.

{\color{black}
This squared-monopole factor introduces a double pole with the same kinematic
denominator as the exchange propagator, with $m_{ex}$ replaced by $\Lambda$.
Since $\Lambda>0$, the common contour rotation discussed above does not cross
this pole either.
}

For the purpose of comparing the three equations, we retain only the leading temporal term in the vector-exchange numerator and neglect the remaining momentum-dependent terms. This defines the vector-exchange kernel used below and is not intended as a complete covariant treatment of the $D\bar{D}$ interaction.

Ref.~\cite{Wang:2021aql} uses $\beta=-0.90$, $g_V=5.8311$, and $g_\sigma=-0.76$. In our convention, these parameters give the vector strength $\alpha_V=\beta^2g_V^2/(4\pi)=2.191$ and the scalar strength $\alpha_\sigma=g_\sigma^2/(4\pi)=0.046$. For the isoscalar channel, the $\rho$- and $\omega$-exchange terms have weights $3/4$ and $1/4$, respectively; hence, $\alpha_\rho=3\alpha_V/4$ and $\alpha_\omega=\alpha_V/4$ in Eq.~(\ref{equation:BS-potential-dipole}). The signs of $\beta$ and $g_\sigma$ do not change these squared effective strengths.

\begin{figure}[htbp]
\centering
\includegraphics[scale=0.75]{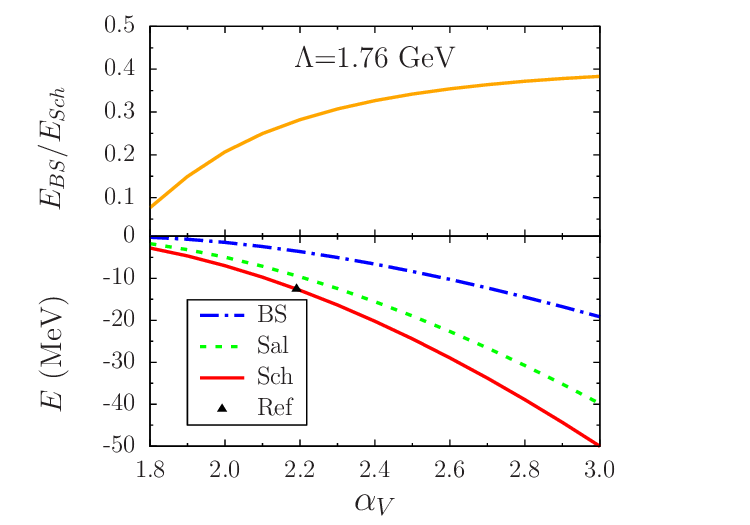}
\caption{The ground-state binding energy $E$ of the $D\bar{D}$ system as a function of $\alpha_V$. Here, $\Lambda=1.76$ GeV, and all other parameters are fixed at the values given in the text. The notations are the same as those in Fig.~\ref{Fig:Eb-alpha-m=1}, and the result from Ref.~\cite{Wang:2021aql} is also presented.}
\label{Fig:result-DD-lambda}
\end{figure}

The exchanged-boson masses are fixed at $m_{\rho}=0.776$ GeV, $m_{\omega}=0.783$ GeV, and $m_{\sigma}=0.6$ GeV. The constituent masses are set to $m_{D}=m_{\bar{D}}=1.867$ GeV, and the cutoff is taken as $\Lambda=1.46$ or $1.76$ GeV. With these inputs, we report the corresponding binding energies in Tab.~\ref{Tab:result-DD-lambda}. Figure~\ref{Fig:result-DD-lambda} shows the dependence of the binding energy $E$ on $\alpha_V$, with all other parameters held fixed.

\renewcommand\tabcolsep{0.3cm}
\renewcommand{\arraystretch}{1.0}
\begin{table}[h!]
\begin{center}
\begin{tabular}{c|cccc}\bottomrule[1pt]
\diagbox[width=2cm,height=1.0cm]{$\Lambda$}{$E$}
&BS &Sal&Sch&Ref.~\cite{Wang:2021aql}\\\hline
1.46&--&-0.11&-0.30&-0.29\\\hline
1.76&-3.51&-9.39&-12.56&-12.55\\
\bottomrule[1pt]
\end{tabular}
\caption{Ground-state binding energies $E$ of the $D\bar{D}$ system calculated using the BS, Salpeter, and Schr{\"o}dinger equations, together with the results from Ref.~\cite{Wang:2021aql}. A dash indicates that no bound-state solution was found. The units of $\Lambda$ and $E$ are GeV and MeV, respectively.}
\label{Tab:result-DD-lambda}
\end{center}
\end{table}

The results in Tab.~\ref{Tab:result-DD-lambda} show that the binding energy $E$ obtained from the BS equation is much closer to zero than that from the Salpeter and Schr\"odinger equations with the same input parameters. As shown in Fig.~\ref{Fig:result-DD-lambda}, the difference in binding energy increases progressively as $\alpha_V$ grows, with the exchanged-meson masses and the cutoff $\Lambda$ held fixed. Specifically, over the range where the Schr\"odinger binding energy is approximately $1\sim 50$ MeV, the magnitude of the BS binding energy is only $10\% \sim 30\%$ of the Schr\"odinger result. This quantifies the three-dimensional reduction effect for strong coupling to heavy $\sigma$, $\rho$, and $\omega$ exchange at fixed input parameters. Our Schr\"odinger results agree with Ref.~\cite{Wang:2021aql} because we use the same NR parameter set. This agreement alone does not justify transferring those parameters directly to a ladder BS calculation.

For comparison, the inverse binding momentum $R_0$ is about $4.3$~fm for the deuteron with a binding energy of $2.2$~MeV and about $4.6$~fm for a $D\bar{D}$ state with a binding energy of $1$~MeV. These characteristic lengths are similar, but this alone does not make their fixed-parameter OBE reductions equivalent. In chiral nuclear EFT, long-range one-pion exchange is supplemented by short-range interactions fitted to nucleon--nucleon data~\cite{Hammer:2019poc}. The $D\bar{D}$ inputs used here instead follow Ref.~\cite{Wang:2021aql}; they involve strong couplings to heavier exchanged mesons and have not been matched to the ladder BS dynamics. Therefore, the reliability of a nonrelativistic description of the deuteron does not conflict with the large ladder BS--Schr\"odinger difference found here at fixed OBE parameters. A systematic matching for the hadronic OBE model used here is beyond the scope of the present work.

\section{Summary}

{\color{black}
In summary, the large numerical differences reported here quantify a fixed-kernel three-dimensional reduction effect within the ladder OBE model. They should not be read as complete predictions for a physical hadronic molecule: nearby coupled channels and corrections to the irreducible kernel must be assessed separately. The comparison nevertheless shows that OBE parameters used in a reduced Schr\"odinger description cannot in general be transferred unchanged to a four-dimensional ladder BS calculation without matching.
}

\section*{Acknowledgements}
H. Q. Zhou acknowledges Yong-Hui Lin, Dian-Yong Chen and Zhi-Yong Zhou for their helpful discussions. This work was funded by the National Natural Science Foundation of China (NSFC) under Grants Nos.~12075058 and 12150013 and supported by the Postgraduate Research
\& Practice Innovation Program of Jiangsu Province under Grants
No.~KYCX25\_0420.

\bibliography{ref}

\end{document}